\documentstyle[epsfig]{aipproc}

\begin{document}

\begin{flushright}
SLAC-PUB-8551\\
August 2000
\end{flushright}

\title{$b \to s \ell^+ \ell^-$ decays in and beyond the Standard Model 
\thanks{Work supported by the Department of Energy, 
Contract DE-AC03-76SF00515} }

\author{Gudrun Hiller$^*$}
\address{$^*$Stanford Linear Accelerator Center, 
              Stanford University, Stanford, CA 94309, USA }

\maketitle

\begin{abstract}
We briefly review the status of rare radiative and semileptonic
$b \to s (\gamma, \ell^+ \ell^-)$, ($\ell=e,\mu$) decays.
We discuss possible signatures of new physics in these modes and
emphazise the role of the exclusive channels.
In particular, measurements of the Forward-Backward asymmetry in 
$B \to K^* \ell^+ \ell^-$ decays and its zero provide a clean test of
the Standard Model, complementary to studies in $b \to s \gamma$ decays.
Further, the Forward-Backward CP asymmetry in 
$B \to K^* \ell^+ \ell^-$ 
decays is sensitive to possible non-standard
sources of CP violation mediated by flavor changing neutral current 
$Z$-penguins.

\end{abstract}

\section*{Introduction}

Flavor changing neutral current (FCNC) $b$ decays do not occur at tree
level in the Standard Model (SM).
Being loop induced, they feel scales of order ${\cal{O}}(m_W,m_t)$
and in principle much
higher ones, making them important probes of the flavor sector of the
SM and beyond.

Rare radiative $b \to s \gamma$ decays proceed via
so-called electromagnetic penguins. They have been measured in
exclusive $B \to K^* \gamma$ \cite{cleo99excl} and 
inclusive $B \to X_s \gamma$ \cite{cleobsg,lepbsg} decays.
In dilepton channels $b \to s \ell^+ \ell^-$ ($\ell=e,\mu$), 
we identify two additional structures in the Feynman
diagrams: boxes and $Z$-penguins.
None of the dilepton modes has been detected to date, but we expect
large data samples from operating $B$-factories (CLEO,BaBar,Belle), dedicated
$B$-physics programmes at colliders (Tevatron Run II,Hera-B) and LHC-B in
the long term.
Corresponding $b \to d$ transition amplitudes are CKM suppressed
$V_{td}^*/V_{ts}^* \propto \lambda \sim 0.22$.

The existing best bound in the dimuon channels for inclusive decays is
${\cal{B}}(B \to X_s \mu^{+} \mu^{-})  < 5.8 \cdot 10^{-5}$ at 90\%
C.L. \cite{CLEO98},
which is one order of magnitude above the NLO SM expectation
${\cal{B}}(B \to X_s \mu^{+} \mu^{-})_{SM} = 5.7 \pm 1.1 \cdot 10^{-6}$
\cite{AHHM}. Note that the NNLO calculation in $b \to s \ell^+ \ell^-$ is only
partially available \cite{bobethetal}.
Corresponding bounds for the exclusive channels are
${\cal{B}}(B^+ \to K^{+} \mu^{+} \mu^{-}) < 5.2 \cdot 10^{-6}$,
${\cal{B}}(B^0 \to K^{*0} \mu^{+} \mu^{-}) < 4.0 \cdot 10^{-6}$
at 90\% C.L. \cite{CDF99}
and their respective SM predictions are
${\cal{B}}(B \to K \mu^{+} \mu^{-})_{SM}=5.9 \pm 2.1 \cdot 10^{-7}$,
${\cal{B}}(B \to K^* \mu^{+} \mu^{-})_{SM}=2.0 \pm 0.7 \cdot 10^{-6}$
with the dominant theoretical uncertainty resulting from hadronic matrix
elements, which are estimated here using
Light cone sum rules \cite{ABHH99}.
Currently, the exclusive $B \to K^* \mu^{+} \mu^{-}$ decay has the most
interesting bound, which is only a factor of 2 away from the SM
prediction. Despite larger theoretical uncertainty than in the inclusive
cases, rare exclusive decays are more accessible experimentally in the
near future and have observables (e.g.~existence and 
position of the zero of the
Forward-Backward asymmetry discussed below),
which are as clean as the respective ones in the inclusive modes.

\section*{Model independent analysis \\photon and Z penguins}

The calculational tool for the description of $b\to s (\gamma,
\ell^+ \ell^-)$ decays is the low energy effective Hamiltonian
${\cal{H}}_{eff} \sim G_F V_{t s}^\ast  V_{tb} \sum_{i=1}^{10} C_i(\mu)
O_i(\mu)$ \cite{heff}.
This enables the analysis of relevant observables in a model
independent way, with the goal being to extract the Wilson coefficients $C_i$
from data \cite{agm95}.
The mayor player is the $bs\gamma$ vertex 
$O_7 \sim m_b \bar{s}_L \sigma _{\mu \nu} b_R F^{\mu \nu}$.
Its effective coupling strength $C_7^{eff}$ is related
to the branching ratio ${\cal{B}}(B \to X_s \gamma) 
\sim |C_7^{eff}|^2$ thus
$0.25 \leq |C_7^{eff}| \leq 0.37$ \cite{cleobsg,ABHH99}
which is in good agreement with the SM value $C_7^{eff}|_{SM}=-0.31$
at $\mu=m_b$ at leading log.
We see that the $b \to s \gamma$ data fix the modulus of $C_7^{eff}$, but not
its sign (phase).

In $b \to s \ell^+ \ell^-$ decays, in addition to $O_7$, also 4-Fermi
operators involving dileptons contribute, given by
$O_9 \sim \bar{s}_L \gamma^{\mu}  b_L\bar{\ell} \gamma_{\mu} \ell$,
$O_{10} \sim \bar{s}_L \gamma^{\mu} b_L \bar{\ell} \gamma_{\mu}\gamma_5 \ell$.
Due to the charge assignments of lepton-$Z$ couplings the $Z$-penguin
contribution to $C_9$ is suppressed with respect to $C_{10}$ by
$(\bar{\ell} \ell Z|_V)/({\bar{\ell} \ell Z|_A})=-1+4 \sin^2\theta_W
\sim -0.08$. We thus identify $C_{10}$ as a measure of the $sZb$ coupling
modulo the box contribution \cite{GGG}.

Decomposition of the $B \to K^{*} \mu^{+} \mu^{-}$ branching ratio yields
${\cal{B}} =a |C_7^{eff}|^2 +b |C_9|^2+c |C_{10}|^2 
+d C_7^{eff} C_9 +e C_7^{eff} +f C_9 +g$ \cite{ABHH99}.
Using the CDF bound \cite{CDF99} on this mode and allowing $C_7^{eff}$
to have both SM-like and SM-opposite signs gives the present best
bound on the strength
of generic FCNC $Z$-penguins of $|C_{10}| \leq 10$,
which is a 
factor of 2-3 larger than the SM value $C_{10}|_{SM}=-4.7$ \cite{GGG}.
Scenarios with non-standard $Z$-penguins arise in many extensions of
the SM like such as supersymmetry, 4th generation and $Z'$ \cite{GGG}.
Another interesting possibility to test the $sZb$ vertex arises in
$b \to s \nu \bar{\nu}$ decays, since here no photon penguins
contribute.

\section*{Exclusive $B \to K,K^* \ell^+ \ell^-$ decays}

\begin{figure}[b!] 
\centerline{\epsfig{file=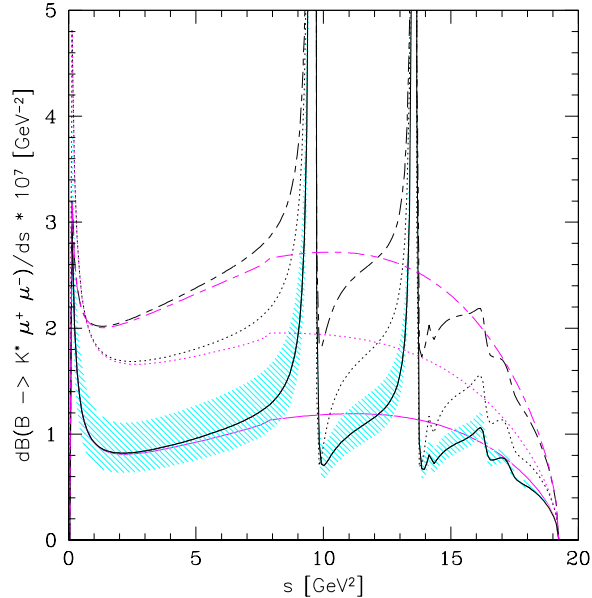,height=3.5in,width=3.5in}}
\vspace{10pt}
\caption{Dilepton invariant mass spectrum in
$B \to K^* \mu^+ \mu^-$ decays. 
Figure taken from \protect\cite{ABHH99}.}
\label{fig1}
\end{figure}

Supersymmetric effects in inclusive $b \to s \ell^+ \ell^-$ decays have been
studied in \cite{susy,goto98}.
The reach of a new physics search in the dilepton invariant mass
distribution in $B \to K^* \mu^+ \mu^-$ decays is exemplified in
Fig.~\ref{fig1} \cite{ABHH99}. Supergravity (SUGRA) (dotted) and a
supersymmetric scenario with non-minimal sources of flavor violation
in the mass insertion approximation (dashed), can be well
discriminated
from the SM (solid) and its hadronic uncertainties (shaded area);
the upper curves contain resonant $c \bar{c}$ background via
$B \to K^* \Psi^{i} \to K^* \mu^+ \mu^-$, lower ones are
pure short-distance contributions \cite{ccbar}.
Note that the dashed curve saturates the experimental bound in this channel.
Similar findings are valid for $B \to K \mu^+ \mu^-$ decays, which
however show less sensitivity to $C_7^{eff}$ as the photon pole at
$s=0$ is absent.
In either case $C_7^{eff}>0$ (opposite-to-SM sign) enhances the rates through
constructive interference of $C_7^{eff}$ with $C_9$.
\subsubsection*{Forward-Backward Asymmetry}

\begin{figure}[b!] 
\centerline{\epsfig{file=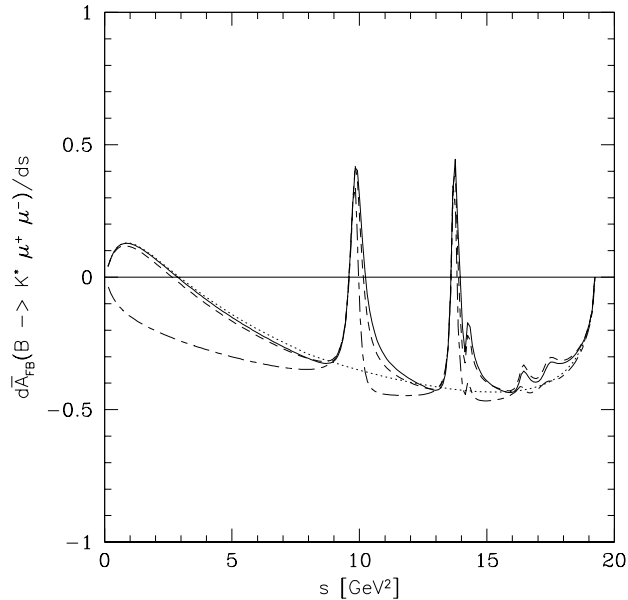,height=3.5in,width=3.5in}}
\vspace{10pt}
\caption{Forward-backward asymmetry for
$B \to K^* \mu^+\mu^-$.
Figure taken from \protect\cite{ABHH99}.}
\label{fig2}
\end{figure}

The Forward-Backward asymmetry in $B \to K^* \ell^+ \ell^-$ decays results
from V/A interference in the lepton pair
$A(s) \sim C_{10} (C_7^{eff}+ \beta (s) Re(C_9^{eff}))$,
shown in Fig.~\ref{fig2} \cite{ABHH99}; see \cite{GGG} for a discussion
of the sign of $A$, which is opposite to the Forward-Backward
asymmetry $\bar{A}$ of the CP conjugate channel in the CP conserving limit.
In the SM ($C_7^{eff}<0$, solid curve), $A$ has a zero around
$s_0\sim 3 \mbox{GeV}^2$, which would disappear if $C_7^{eff}$ would
have the opposite sign (long-short-dashed curve). 
The existence of a zero in the Forward-Backward asymmetry 
in $B \to K^* \ell^+ \ell^-$ decays below the $J/\Psi$ resonance is an 
important test of the SM and is independent of
hadronic matrix elements \cite{ABHH99}. Positive 
$C_7^{eff}$ occurs generically in supersymmetric theories
\cite{susy}, 
but only for large $tan \beta$ in relaxed and/or minimal SUGRA \cite{goto98}.

Further, the position of the zero $s_0$ has very small hadronic uncertainties 
\cite{burdman}. In the limit where the final hadron has large energy,
i.e., small dilepton mass, the Large Energy Effective Theory \cite{LEET} is 
applicable and here all form factors cancel out in the ratios which determines
$s_0$ \cite{ABHH99}.

A recently proposed observable, the  Forward-Backward CP asymmetry
$FB_{CP}$ in
$B \to K^* \ell^+ \ell^-$ decays probes the phase of $C_{10}$ or of the
$sZb$ vertex \cite{GGG}, respectively.
Defined as 
$FB_{CP} \equiv(A+\bar{A})/(A-\bar{A})=Im C_{10}/Re C_{10} Im C_9^{eff}/Re
C_9^{eff} (1+\dots)$, its magnitude scales with $Im C_9^{eff}=Im Y$,
where
$Y(s)$ contains contributions from resonant and
non-resonant $c\bar{c}$ states \cite{ccbar}; 
$ImY=0$ below threshold, so it is sizeable
only in the high dilepton mass region. Integration over
$m_{\Psi'}^2 < s \leq (m_B-m_{K*})^2$ yields
$\Delta FB_{CP}=(3 \pm 1) \% Im C_{10}/Re C_{10}$, which can be large
in the case of ${\cal{O}}(1)$ phases.
Hadronic uncertainties are not small, but the SM background is below $10^{-3}$
and any effect above this would be due to
a non-SM source of CP violation \cite{GGG}.

\section*{Radiative rare $B$-decays}

The SM branching ratio in $B \to X_s \gamma$ decays is known at NLO with
10 \% accuracy 
${\cal{B}}(B \to X_s \gamma)=(3.32 \pm^{0.00}_{0.11} \pm^{0.00}_{0.08}
\pm^{0.26}_{0.25}) \cdot 10^{-4}$
\cite{greubborzumati}.
This is in good agreement with data from LEP
${\cal{B}}(B \to X_s \gamma) =(3.11 \pm 0.80\pm 0.72) \cdot 10^{-4}$ 
\cite{lepbsg}, and CLEO,
${\cal{B}}(B \to X_s \gamma)=(3.15 \pm 0.35 \pm 0.32 \pm 0.26) \cdot 10^{-4}$
\cite{cleobsg}.

A promising observable in $b \to s \gamma$ decays where dramatic
signals of possible physics beyond the SM
could show up is the CP asymmetry in the rate
$a_{CP} \equiv
(\Gamma(\bar{B} \to X_s \gamma)-\Gamma(B \to X_{\bar{s}} \gamma))/
(\Gamma(\bar{B}\to X_s \gamma)+\Gamma(B \to X_{\bar{s}} \gamma))$ 
\cite{bsgCPXs}. It is very tiny in the SM since
$a_{CP} \propto \alpha_s(m_b)\eta \lambda^2$, 
$\eta, \lambda$ are Wolfenstein parameters, thus
$a_{CP}\leq 1 \%$ \cite{bsgCPXs}.
However, large effects of (10-50) \% are possible in scenarios with
an enhanced chromo-magnetic
dipole operator $C_8$ in the $b s \mbox{g}$ vertex.
CP asymmetry in exclusive $B \to K^* \gamma$ decays has a less clean
prediction due to strong phases, however, in the SM,
$a_{CP}^{B \to K^* \gamma} \leq {\cal{O}}(1 \%)$ holds \cite{bsgCPKst}.
Any significant deviation from this would signal new physics.
In both inclusive $-0.09 < a_{CP}^{B \to X_s \gamma} < 0.42 $ \cite{cleobsg}
and exclusive cases 
$ a_{CP}^{B \to K^* \gamma} = (8 \pm 13 \pm 3) \% $ \cite{cleo99excl}, the
measurements are not conclusive yet.

\section*{Summary}

Theoretically clear signatures of possible new physics
can be experimentally isolated in $b \to s \gamma$ and $b \to s \ell^+ \ell^-$
decays in the near future via the 
observables ($A(s_0),FB_{CP},a_{CP}$).
It is exciting to see whether the SM passes this next round of FCNC tests.

\end{document}